\documentclass[11pt,twoside]{article}
\usepackage{./asp2010}

\resetcounters

\begin{document}

\title{The Dynamical Properties of Virgo Cluster Disk Galaxies}
\author{Nathalie~N.Q.~Ouellette,$^1$ St\'{e}phane~Courteau,$^1$ Jon~A.~Holtzman,$^2$ Julianne J. Dalcanton,$^3$ Michael~McDonald,$^4$ Yucong~Zhu$^5$
\affil{$^1$Department of Physics, Engineering Physics and Astronomy, Queen's University, Kingston, ON, K7L 3N6, Canada}
\affil{$^2$Department of Astronomy, New Mexico State University, Las Cruces, NM, 88003-8001, USA}
\affil{$^3$Department of Astronomy, University of Washington, Seattle, WA, 98195-1580, USA}}
\affil{$^4$Kavli Institute for Astrophysics and Space Research, Massachusetts Institute of Technology, Cambridge, MA, 02139, USA }
\affil{$^5$Center for Astrophysics, Harvard University, Cambridge, MA, 02138, USA}

\begin{abstract}
By virtue of its proximity, the Virgo Cluster is an ideal laboratory for testing our understanding structure formation in the Universe. In this spirit, we present a dynamical study Virgo galaxies as part of the Spectroscopic and H-band Imaging of Virgo (SHIVir) survey. H$\alpha$ rotation curves (RC) for our gas-rich galaxies were modelled with a multi-parameter fit function from which various velocity measurements were inferred. Our study takes advantage of archival and our own new data as we aim to compile the largest Tully-Fisher relation (TFR) for a cluster to date. Extended velocity dispersion profiles (VDP) are integrated over varying aperture sizes to extract representative velocity dispersions (VDs) for gas-poor galaxies. Considering the lack of a common standard for the measurement of a fiducial galaxy VD in the literature, we rectify this situation by determining the radius at which the measured VD yields the tightest Fundamental Plane (FP). We found that radius to be at least 1 $R_{\rm e}$, which exceeds the extent of most dispersion profiles in other works.
\end{abstract}

\section{Introduction}

The mechanisms by which cluster substructure formed remain somewhat open-ended despite decades of observations and theoretical modeling. Whether galaxies came to be via secular evolution, hierarchical formation or some combination of the two is still under investigation. Studying the internal dynamics of cluster galaxies can give vital clues to the processes involved in their evolution. Ingrained in each velocity measurement is the history of that galaxy: tidal interactions, bar formation, mergers, etc.

A massive undertaking was initiated by \citet{Binggeli1985} to image an area of 140 deg$^{2}$ around the Virgo core in blue wavelengths: the Virgo Cluster Catalog (VCC). Through our own SHIVir survey, we have compiled existing and acquired new photometric and spectroscopic observations of Virgo galaxies \citep{McDonald2009,McDonald2011,Roediger2011a,Roediger2011b}. The present work aims to fold in spectroscopically determined internal stellar dynamics to our spatially resolved observations of Virgo galaxies. Examination of extended dynamical profiles should reveal features which correlate with overall galaxy physical parameters.

Our main goals are twofold. First, extract and study the spatially resolved VDPs of gas-poor galaxies in the Virgo Cluster. Looking deeply into these galaxies should reveal extended dynamical profiles whose shapes may, in turn, correlate with other galaxy physical parameters. Measurements of galaxy structural parameters also call for great uniformity in their definition in order to avoid interpretation bias. VDs are indeed fraught with confusion given their many heterogeneous definitions in the literature. VDs are often measured at a varying fraction of an effective radius, or at some pre-determined fixed physical radius.  \citet{Jorgensen1996} established such a system normalised to $R_{\rm e}/8$, but it samples dispersion profiles in a regime where significant variations exist even for galaxies of a comparable mass. Second, we wish to apply a similar analysis for gas-rich galaxies. That is, first confirm the best measure of line-of-sight velocity based on galaxy emission lines and apply this definition to our SHIVir galaxies in order to derive tight scaling relations. Of ultimate interest for the connection of observations with theoretical models of structure formation is to combine both types of velocity measures for all galaxies to produce a complete galaxy velocity/mass function \citep{Dutton2011,TrujilloGomez2011}.

\section{Survey \& Data}

For the purpose of SHIVir investigations, stringent selection criteria \citep{McDonald2011} were applied to the VCC. These criteria reduce the VCC to a magnitude-limited catalog of 286 SHIVir galaxies for which we have collected $H$-band photometry \citep{McDonald2011}.

Our long-slit spectroscopic observations were collected on the 3.5m telescope at the Apache Point Observatory. The IDL routine \texttt{pPXF} \citep{Cappellari2004} was then used to extract RCs and VDPs.  RCs from H$\alpha$ emission lines were extracted for 34 SHIVir late-type galaxies. Conversely, integrated VDs and RCs from absorption features were obtained for 31 SHIVir early-type galaxies (ETG). Due to weather, time and telescope size constraints, our spectroscopic catalog consists mostly of the brightest and most massive SHIVir galaxies. The fainter ones will soon be observed on larger ($>$8m) telescopes.

\articlefiguretwo{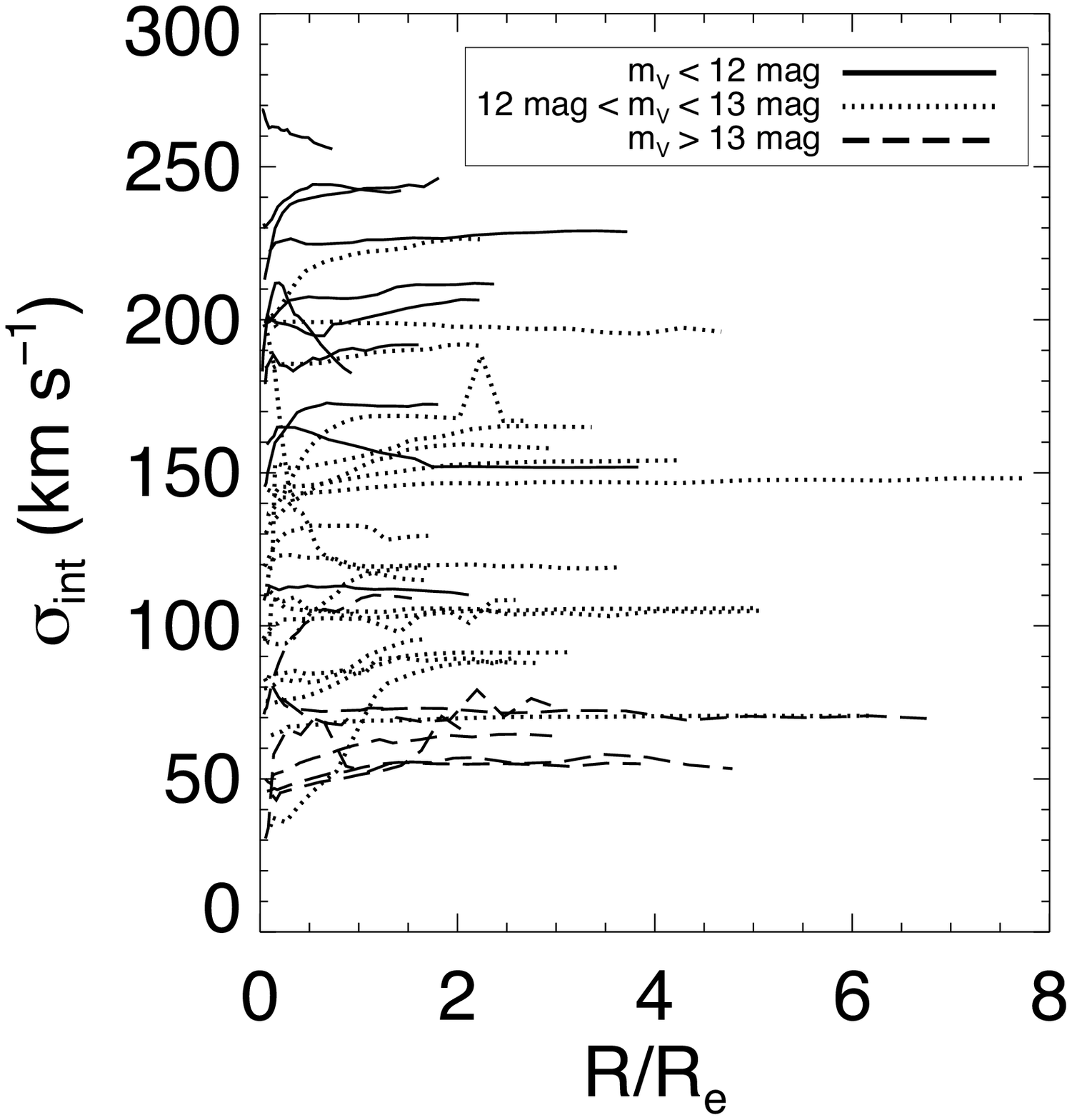}{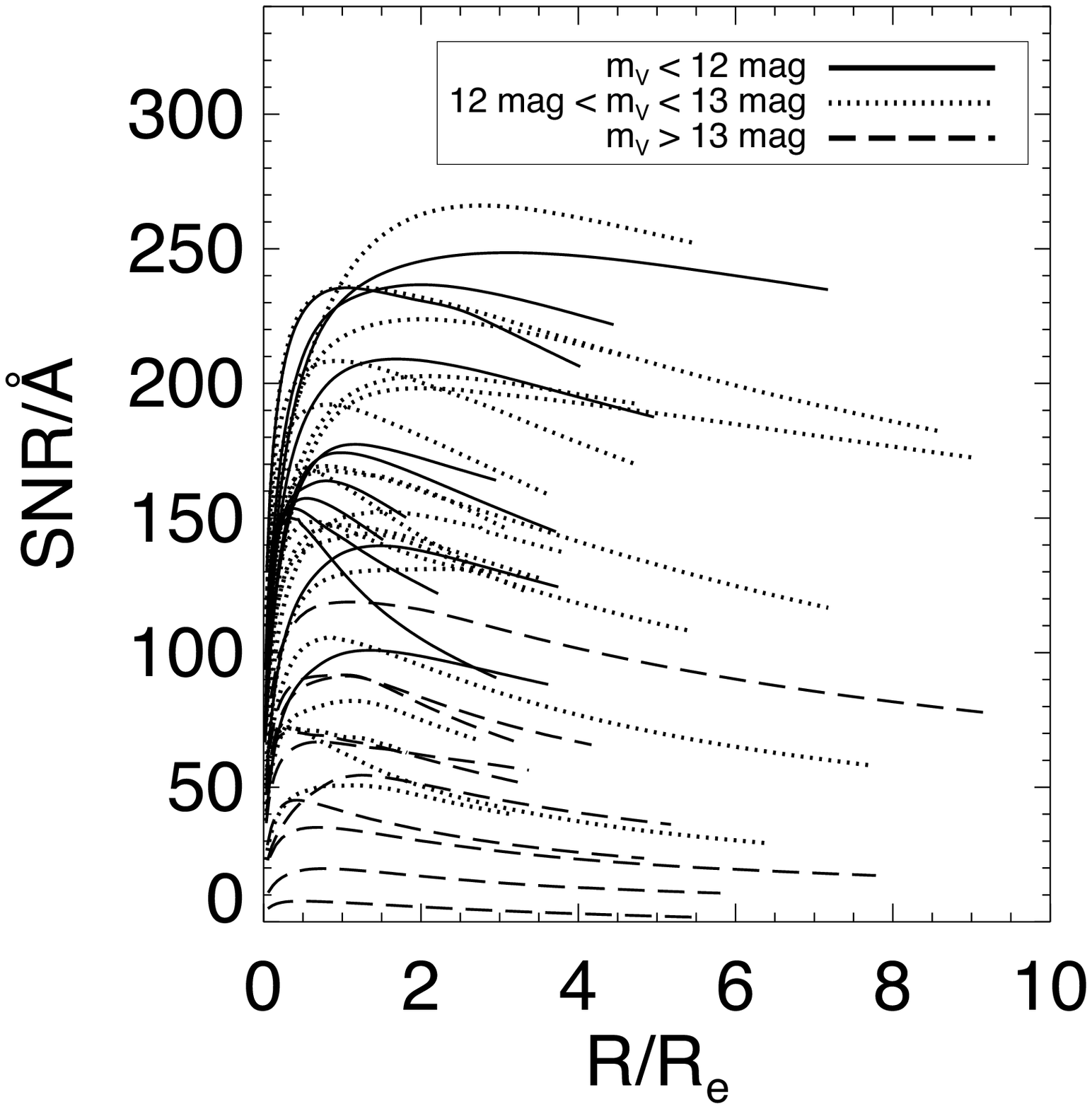}{fig}{(Left) Integrated dispersion profiles as a function of aperture size for 31 SHIVir ETGs; (Right) SNR growth curves associated with the values on the right for the same aperture range.}

\section{Results}

\subsection{Dynamical Profiles}

While our SHIVir H$\alpha$ RCs match nicely with those of \citet{Rubin1997} and \citet{Chemin2006} [for 17 overlap], they typically do not extend as far as those of field galaxies of similar stellar masses \citep{Courteau1997}. Many of them even fail to reach 1 $R_{\rm e}$. They were also compared to the VIVA survey H{\footnotesize I} RCs \citep{Chung2009}. We expect these to be considerably more extended than our own RCs due to the large size of a typical galaxy's H{\footnotesize I} gas disk. However, many of these H{\footnotesize I} RCs were also found to be truncated: some were even less extended than our own H$\alpha$ curves. Truncation of both the stellar and H{\footnotesize I} gas disks appears to be due to environmental effects of the cluster. Comparisons between our H$\alpha$ RCs and those determined from absorption lines indicate that, for some of our disk galaxies, the gas is rotating at slighter higher velocities than the stars: no more than a difference of 15\%, as expected from asymmetric drift of the stars. The exception is VCC1003 which contains a LINER that is most likely causing additional turbulence in the gas. Rotational velocities taken from our H$\alpha$ RCs have been used to populate our TFR and to study the variations of galaxy RCs with structural parameters (Ouellette et~al., in prep.)

Our VDPs extend to 1--4 $R_{\rm e}$, a somewhat larger range than found for similar ETG studies, e.g. SAURON/ATLAS3D \citep{Bacon2001,Cappellari2011}. Dynamical surveys based on external tracers (e.g. PNe, globular clusters) can reach even further but such data sets are currently very limited due to the required long integrations \citep{Courteau2013}. Our radial sampling (1--4 $R_{\rm e}$) probes the transition from baryonic to dark matter dominance. Most profile shapes are also either peaked or depressed (so called ``sigma drop" galaxy) at the centre. Such sigma-drop systems are found to be mostly associated with ring features, in agreement with \citet{Comeron2010}. We also find them to be most prevalent in S0 galaxies. As we increase the size of our catalogue and probe intermediate and low-mass regimes, we aim to solidify the links between VDP shape and other structural parameters. Our findings are contrasted against those for field galaxies in order to obtain a complete evolutionary framework for cluster galaxies.

\subsection{Integrated Velocity Dispersions}

As mentioned earlier, extraction of representative VDs, e.g. for the computation of dynamical mass, can be confusing in light of the many different definitions in the literature. A possible approach to selecting an ideal VD measure is to choose the one which minimises a fundamental correlation in ETGs. Rather than measure a VD at a single aperture size, we study the integrated VD within increasing aperture sizes (Fig.~\ref{fig} (left)) and identify the optimal aperture which minimises the FP. VDs have historically been measured within fractions of $R_{\rm e}$ or at the centre ($\sigma_0$). However, VDPs vary significantly (by as much as 20\%) within 1$R_{\rm e}$, emphasizing the importance of aperture size in dynamical studies. Our tests indicate that a VD measurement at 1$R_{\rm e}$ or beyond minimises the FP scatter (see \citet{Cappellari2013}; although, their use of small IFUs meant that they often had to extrapolate their VDPs to 1$R_{\rm e}$). We find the peak signal-to-noise ratio (SNR) to be typically reached between 0.5 and 2 $R_{\rm e}$ (Fig.~\ref{fig}, right), which means that a VD measured within such an aperture optimises flux.

\section{Future Work}

We have exhausted observing possibilities on 4~m-class telescopes and will be using 8--10m-class telescopes to extend our analysis into intermediate- and low-mass galaxies. We have already investigated the effect of aperture size within which VD is measured on the scatter of scaling relations. We intend to perform a principal component analysis on the many galaxy structural parameters for all SHIVir galaxies in order to determine which metric controls the FP and its scatter. Our circular velocity function for the Virgo Cluster will enable direct tests of theoretical structure formation models \citep{Miller2013}.

\acknowledgements Thanks to Marc Seigar and his team for organizing a wonderful conference. NNQO and SC acknowledge financial support from NSERC.

\bibliography{author}

\end{document}